\documentclass[fleqn,usenatbib]{mnras}

\renewcommand{\d}[1]{\ensuremath{\operatorname{d}\!{#1}}}
\usepackage{newtxtext,newtxmath}

\usepackage[T1]{fontenc}

\DeclareRobustCommand{\VAN}[3]{#2}
\let\VANthebibliography\thebibliography
\def\thebibliography{\DeclareRobustCommand{\VAN}[3]{##3}\VANthebibliography}

\usepackage{graphicx}	
\usepackage{amsmath}	
\usepackage{listings}
\usepackage{multicol}

\title[Tidal dissipation in planets]{Tidal dissipation in rotating and evolving giant planets with application to exoplanet systems}

\author[Lazovik et al.]{
Yaroslav A. Lazovik,$^{1,2}$\thanks{E-mail: yaroslav.lazovik@gmail.com}
Adrian J. Barker,$^{3}$\thanks{E-mail: A.J.Barker@leeds.ac.uk} 
Nils B. de Vries$^{3}$\thanks{E-mail: mmnbdv@leeds.ac.uk} and
Aurélie Astoul$^{3}$\thanks{E-mail: A.A.V.Astoul@leeds.ac.uk}
\\
$^{1}$Lomonosov Moscow State University, Faculty of Physics, 1 Leninskie Gory, bldg.2, Moscow, 119991, Russia\\
$^{2}$Sternberg Astronomical Institute, Lomonosov Moscow State University, Universitetsky pr. 13, Moscow, 119234, Russia \\
$^{3}$ School of Mathematics, University of Leeds, Leeds LS2 9JT, UK
}

\date{Accepted 2023 November 26. Received 2023 November 24; in original form 2023 August 07}

\pubyear{2023}

\begin{document}
\label{firstpage}
\pagerange{\pageref{firstpage}--\pageref{lastpage}}
\maketitle

\begin{abstract}
We study tidal dissipation in models of rotating giant planets with masses in the range $0.1 - 10 M_\mathrm{J}$ throughout their evolution. Our models incorporate a frequency-dependent turbulent effective viscosity acting on equilibrium tides (including its modification by rapid rotation consistent with hydrodynamical simulations) and inertial waves in convection zones, and internal gravity waves in the thin radiative atmospheres. We consider a range of planetary evolutionary models for various masses and strengths of stellar instellation. Dissipation of inertial waves is computed using a frequency-averaged formalism fully accounting for planetary structures. Dissipation of gravity waves in the radiation zone is computed assuming these waves are launched adiabatically and are subsequently fully damped (by wave breaking/radiative damping). We compute modified tidal quality factors $Q'$ and evolutionary timescales for these planets as a function of their ages. We find inertial waves to be the dominant mechanism of tidal dissipation in giant planets whenever they are excited. Their excitation requires the tidal period ($P_\mathrm{tide}$) to be longer than half the planetary rotation ($P_\mathrm{rot}/2$), and we predict inertial waves to provide a typical $Q'\sim 10^3 (P_\mathrm{rot}/1 \mathrm{d})^2$, with values between $10^5$ and $10^6$ for a 10-day period. We show correlations of observed exoplanet eccentricities with tidal circularisation timescale predictions, highlighting the key role of planetary tides. A major uncertainty in planetary models is the role of stably-stratified layers resulting from compositional gradients, which we do not account for here, but which could modify predictions for tidal dissipation rates.
\end{abstract}




\section{Introduction} \label{intro}
Tidal interactions play a major role in the dynamics of star-planet and stellar binary systems, leading to planetary orbital migration \citep[e.g.][]{Jackson1,Jackson2,Villaver,Bolmont2016,Gallet,Bolmont2,Ahuir,Lazovik}, orbital circularization \citep[e.g.][]{Witte,Nagasawa,Beauge,Barker4}, spin-orbit re-alignment \citep[e.g.][]{Barker1,Winn,Lai,Hamer}, and rotational evolution \citep[e.g.][]{Bolmont1,Penev,Gallet1,Gallet2,Barker4}. Indeed, tidal interactions alter the architectures of exoplanetary systems, and the rotations of stars and planets. However, theoretical predictions for tidal evolution are currently uncertain, and they highly depend on the specific prescriptions employed in the aforementioned papers. This motivates us to work towards developing more realistic treatments of tidal flows in stars and planets.

Some early steps towards a theory of tides were made long before the discovery of the first exoplanet from studying the Earth-Moon system  
\citep{Darwin}. More than a hundred years of research has substantially refined our knowledge of tidal interactions in fluid bodies such as stars and giant planets. Motivated by binary stars, some crucial steps were made by \cite{Zahn1, Zahn2, Zahn3}, who separated the contributions of equilibrium (non-wavelike) and dynamical (wavelike) tides in linear theory. Dynamical tides can be further divided into inertial waves (hereafter IWs, or magneto-inertial) and internal gravity waves (hereafter GWs, or magneto-gravito-inertial), propagating in the convective and radiative regions, respectively \citep[e.g.][]{T1998,GD1998,Ogilvie1,Wu2005b,OL2007,GL2009,Weinberg2012,IvPapCh2013,LO2018}. Recently, \citet[][hereafter B20]{Barker3} applied the latest tidal theory to compute modified tidal quality factors $Q'$ (essentially the ratio of the maximum tidal energy stored to the energy dissipated in one tidal period -- a quantity essential for tidal modelling) and tidal evolutionary timescales in a range of stellar models, developing prescriptions for the dissipation of tides of various types (i.e.~equilibrium tides, IWs and GWs). These were implemented in \cite{Lazovik,Lazovik1} to explore the secular evolution of hot Jupiter systems over a wide parameter space, and employed to provide an explanation for close solar-type binary circularization in \cite{Barker4}.

Even after the first exoplanet detection, attention has primarily focused on computing tidal dissipation inside stars, while dissipation in planetary interiors has typically been restricted to Solar System objects. A comprehensive application of tidal theory to rotating planetary models was performed in \cite{Ogilvie1} (see also \citealt{Wu2005b} and \citealt{IP2007}), which opened the doors to a new direction of research. Giant planets share many similarities in structure with stars. They are both primarily multi-layer fluid bodies containing both convective and radiative regions, albeit planets may also have solid cores. But there are important differences: planets tend to rotate faster (compared with both their dynamical and convective frequencies) such that IWs are almost always likely to be important and hot Jupiters also have larger tidal amplitudes than planet-hosting stars, thereby requiring consideration of nonlinear tidal mechanisms like the elliptical instability \citep[that excites IWs in convection zones, e.g.][]{B2016,Nils2023} or other nonlinear IW interactions \citep[e.g.][]{Favier2014,AB2022,AB2023}. The interaction between equilibrium tides and turbulent convection is also expected to be far into the regime of fast tides (where tidal frequencies exceed convective turnover frequencies) because convection is typically much slower in planets than in stars \citep{GN1977,Nils2023}. Hence, the reduction in the turbulent viscosity for fast tides must be accounted for \citep[e.g.][]{GN1977,OL2012,VB2020b,DBJ2020b}. Furthermore, convection is likely to be influenced by rapid rotation, which can modify this interaction \citep[e.g.][]{S1979,Barker2,Mathis2016,CBLB2020,Dandoy2023,Nils2023}.

The role of IWs for planetary tidal dissipation has been explored in prior work \citep[e.g.][]{Ogilvie1,Wu2005a,Wu2005b,O2009,IP2007,GL2009,IP2010,PI2010,O2013,Nils2023}. However, a detailed study of the evolution of tidal dissipation rates and tidal quality factors $Q'$ from equilibrium and dynamical tides following planetary evolution has never been performed previously to our knowledge, though \cite{TM2021} performed computations for just the equilibrium tide, assuming a different mechanism dissipates the tidal flow to what we consider. We use new models of giant planet interiors with masses in the range from $0.1$ to $10 M_\mathrm{J}$ computed with the {\sc MESA} code \citep{MESA1,MESA2,MESA3,MESA4,MESA5}, with various strengths of stellar instellation so as to model both hot and cold planets, to theoretically calculate tidal dissipation rates \citep[thereby extending][for low-mass stars]{Barker3}. In Sec.~\ref{sec:method}, we describe our model. Our results are presented in Sec.~\ref{sec:results} and applied to exoplanet eccentricities in Sec.~\ref{sec:discussion}.

\section{Methods}
\label{sec:method}
\subsection{Tidal dissipation mechanisms}

We consider a giant planet of mass $M_\mathrm{pl}$ and radius $R_\mathrm{pl}$ and employ spherical coordinates centred on the body with radial coordinate $r$. The intensity of tidal dissipation is often quantified by the (modified) tidal quality factor $Q'$, which is proportional to the ratio of the maximum energy stored in the tide to the amount dissipated in one period \citep[e.g.][]{G1963,O2014}. More effective dissipation corresponds to lower $Q'$. Here we consider three mechanisms of tidal dissipation: equilibrium (non-wavelike) tides damped by their interaction with turbulent (rotating) convection, IWs, and GWs. Each mechanism is characterized by its corresponding tidal quality factor ($Q'_\mathrm{eq}$, $Q'_\mathrm{iw}$, and $Q'_\mathrm{gw}$ for equilibrium tides, IWs, and GWs, respectively), and these are calculated within the formalism of B20 (building upon many prior works) with a few modifications that we will describe below. We focus on tides with spherical harmonic degree $l = 2$ and azimuthal wavenumber $m = 2$, which is usually the dominant component in systems with low obliquities. This is likely to be the dominant component of tidal forcing in asynchronously rotating bodies, as well as for eccentricity tides in weakly eccentric but spin-synchronised bodies -- see e.g. Eqs.~4 and 5 in \cite{OL2007} or Table 1 in \cite{O2014}.

The equilibrium tide is a quasi-static fluid response of a perturbed body that is thought to be dissipated through the action of ``turbulent viscosity" in convective zones. The corresponding tidal quality factor is obtained via the expression:
\begin{equation}
    \frac{1}{Q'_\mathrm{eq}} = \frac{16\pi G}{3(2l+1)R_\mathrm{pl}^{2l+1}|A|^{2}}\frac{D_\mathrm{v}}{|\omega_\mathrm{t}|},
    \label{eq:tide_eq1}
\end{equation}
where $G$ is the gravitational constant, $D_\mathrm{v}$ is the rate of viscous dissipation of the equilibrium tide, and $A$ is the amplitude of the tidal potential component (this will not be specified further as it cancels because $D_\mathrm{v}\propto |A|^2$ in linear theory). The tidal forcing frequency is $\omega_\mathrm{t}=2\pi/P_\mathrm{tide}$ (where $P_\mathrm{tide}$ is the tidal period), which is $\omega_\mathrm{t} = 2(n - \Omega_\mathrm{pl})$ for a circular, aligned orbit, where $n$ and $\Omega_\mathrm{pl}$ are the orbital mean motion and planetary spin, respectively. For an eccentric orbit with synchronised\footnote{Perhaps the spin should instead be pseudo-synchronised for an eccentric orbit, but it is not clear that the classical formula of \cite{Hut1981} is valid since it is derived by assuming equilibrium tide damping with a constant time-lag \citep[see also][]{IvPap2004}.} (and aligned) spin we instead have $|\omega_\mathrm{t}|=n$. 

The viscous dissipation rate $D_\mathrm{v}$ is computed using Eqs. (20) to (22) in B20. This quantity depends on the equilibrium tidal displacement vector (defined in section 2 of B20) and the turbulent effective viscosity $\nu_\mathrm{E}$ at each radius. The latter is assumed to be a function of radius and to act like an isotropic shear viscosity, linked to the crude mixing-length theory expectation $\nu_\mathrm{MLT} \propto 
u_\mathrm{c} l_\mathrm{c}$, with $u_\mathrm{c}$ the convective velocity and $l_\mathrm{c}=\alpha H_p$ the mixing-length ($\alpha$ is mixing-length parameter and $H_p$ is pressure scale height). As demonstrated in hydrodynamical simulations \citep[e.g.][]{OL2012,DBJ2020b,VB2020b} and as previously hypothesised using phenomenological arguments \citep[][even if the arguments of both are not supported by simulations, despite agreement regarding the final result with the latter]{Z1966,GN1977}, $\nu_\mathrm{E}$ is reduced for fast tides (where $\omega_\mathrm{t}>\omega_c$, with the convective frequency $\omega_\mathrm{c} = u_\mathrm{c}/l_\mathrm{c}$) in a frequency-dependent manner. This can be accounted for using a piece-wise continuous correction factor depending on the ratio $\omega_\mathrm{t}/\omega_\mathrm{c}$ (for which we employ Eq. (27) of B20, inferred from detailed numerical simulations) at each radius in the planet. Moreover, the rapid rotation expected for giant planets stabilizes convection on large length scales, and a steeper temperature gradient is required to sustain a given heat flux \citep[e.g.][]{S1979,Barker2,CBLB2020}. \cite{Mathis2016} and \cite{Nils2023} have shown that such rapid rotation reduces $\nu_\mathrm{E}$ even further (though interestingly the regime with $\omega_\mathrm{t}\gg \omega_\mathrm{c}$ is not modified by rotation). Following these works, and as confirmed by their simulations, we take into account rapid rotation according to rotating mixing-length theory (RMLT), by multiplying $l_\mathrm{c}$ and $u_\mathrm{c}$ at each radius by $\rm Ro^{3/5}$ and $\rm Ro^{1/5}$, respectively, where $\rm Ro$ is the convective Rossby number ($\mathrm{Ro} = \omega_\mathrm{c}/\Omega_\mathrm{pl}$, based on the non-rotating convective frequency). We demonstrate in Sec.~\ref{sec:results} that the dissipation of equilibrium tides is insufficient to provide significant orbital or spin evolution compared with wavelike tides. 

Motivated by results from the (albeit very idealised), numerical simulations described above, we assume equilibrium tides are damped by their interaction with convection in a way that can be modelled as a local (frequency-dependent) effective viscosity that is positive at each radial location in the planet (and is isotropic for simplicity). Negative values for $\nu_E$ -- corresponding to tidal anti-dissipation -- have been found to occur, particularly at very high tidal frequencies \citep{OL2012,DBJ2020a,VB2020b}, but these are typically negligibly small in magnitude so we neglect their contribution here. On the other hand, we ignore possible contributions to the tidal energy transfer from Reynolds stresses involving tide-tide correlations and gradients of the convective flow \citep[as proposed to be the only important term for fast tides by][even if this interpretation is a drastic over-simplification]{T2021}. This is because we believe that it is not currently possible to estimate contributions from this term without detailed numerical simulations \citep[e.g.][suggesting our overall conclusions regarding the ineffectiveness of equilibrium tides are likely to hold]{BA2021}.
 
Turning to wavelike tides, the tidal quality factor representing inertial wave dissipation is computed following the (low frequency) frequency-averaged formalism of \cite{O2013}, and is calculated according to:
\begin{equation}
    \frac{1}{Q'_\mathrm{iw}} = \frac{32\pi^2 G}{3(2l+1)R_\mathrm{pl}^{2l+1}|A|^{2}}(E_\mathrm{l} + E_\mathrm{l-1} + E_\mathrm{l+1}),
    \label{eq:tide_iw1}
\end{equation}
where the parameters $E_\mathrm{l}$, $E_\mathrm{l-1}$, and $E_\mathrm{l+1}$, are specified by Eqs.~(31)--(33) in B20, fully accounting for the planetary structure. These coefficients are proportional to the squared spin rate, implying that inertial wave dissipation is more efficient in rapidly rotating bodies. Note that $E_\mathrm{l}$ and $E_\mathrm{l\pm 1}$ involve radial integrals that depend to some extent on the assumed core size (inner boundary of the fluid envelope). This dependence on core size is found to be weak in our realistic models though \citep[which is consistent with Fig.~10 of][notably showing compressible polytropic models with indices between 1 and 1.5]{O2013}, unlike results obtained for incompressible models. We adopt impenetrable boundary conditions (with vanishing radial velocity) for inertial waves (\textit{not} the total tide) at the core-envelope boundary and planetary surface, which is appropriate at low frequencies, and follows \citealt{O2013} and \citealt{Barker3}. We do neglect the possibility of very large stably-stratified cores in this work though due to the uncertainties in such planetary models.

This is a simple measure to represent the typical level of dissipation due to inertial waves over the full range of propagation of these waves, which is straightforward to compute in a given planetary or stellar model \citep[e.g.][]{M2015}. Note that this quantity is independent of the specific damping mechanism, and is computed in a model assuming an impulsive encounter to excite all inertial waves, which are then assumed to be subsequently fully dissipated. Modelling tidal evolution of nearly circular or aligned orbits using this quantity involves making assumptions, since this is not rigorously valid \citep[despite its wide usage in this context e.g.][]{Bolmont2016}, but it is believed to be a representative value for inertial wave dissipation. We choose to adopt this approach \citep[following][and many others]{M2015,Bolmont2016,Barker3,Barker4} because this measure is both simpler and much faster to compute (hence amenable to evolutionary studies), and it is also much more robust to the incorporation of additional (or variation in) model physics than the direct linear (or nonlinear) response at a particular frequency.

It should be remembered however that the actual dissipation due to inertial waves at a given $\omega_\mathrm{t}$ could differ substantially from this value \citep[e.g.][]{O2013,AB2022,AB2023}. In particular, predictions for inertial wave dissipation find substantial deviations (potentially by orders of magnitude, either larger or smaller) at a particular frequency -- and between the frequency-averaged measure and the response at a particular frequency -- depending on the degree of density stratification \citep{O2013}, the presence of magnetic fields \citep{LO2018,Wei2018}, differential rotation \citep{BR2013,Guenel2016,AB2022,AB2023}, nonlinearity \citep{Favier2014,B2016,AB2022,AB2023}, convection, varying the microscopic diffusivities \citep{Ogilvie1,O2009}, and the presence of stably-stratified (or different density) inner fluid layers (as opposed to a rigid core) \citep{O2013,Pontin2022,Lin2023,Dewberry2023,Pontin2023b}. However, the frequency-averaged measure has been found to be much more robust regarding the incorporation of magnetic fields \citep{LO2018}, nonlinearity, and to a limited extent differential rotation \citep{AB2023}. It is an open question how reliable this approach will be at modelling a population of individual systems that are each forced at a particular tidal frequency (or range of these) at a given epoch.

We assume that upward propagating gravity waves are excited (adiabatically) at the base of the radiative envelope and are fully damped (e.g.~by radiative diffusion or wave breaking) before propagating back to their launching sites \citep[e.g.][]{Zahn1,Lubow1997,GD1998}. The corresponding tidal quality factor is then given by (B20):
\begin{equation}
    \frac{1}{Q'_\mathrm{gw}} = \frac{2 \left[\Gamma\left(\frac{1}{3}\right)\right]^2}{3^{\frac{1}{3}}(2l+1)(l(l+1))^{\frac{4}{3}}} \frac{R_\mathrm{pl}}{G M_\mathrm{pl}^2} \mathcal{G} |\omega_\mathrm{t}|^{\frac{8}{3}}.
    \label{eq:tide_gw1}
\end{equation}
The quantity $\mathcal{G}$ depends on the planetary conditions at the radiative/convective interface:
\begin{equation}
\mathcal{G} = \sigma_\mathrm{b}^2 \rho_\mathrm{b} r_\mathrm{b}^5 \bigg|\frac{\d \,\mathcal{N}^2}{\d \,\ln\,r}\bigg|_{r=r_\mathrm{b}}^{-\frac{1}{3}}.
    \label{eq:tide_gw2}
\end{equation}
Subscript $\mathrm{b}$ denotes the base of the radiative envelope, so $r_\mathrm{b}$ and $\rho_\mathrm{b}$ are the corresponding radius and density, respectively, $\mathcal{N}$ is the Brunt-V\"ais\"al\"a\ frequency, and the parameter $\sigma_\mathrm{b}$ is determined numerically by the derivative of the dynamical tide radial displacement (see Eq.(43) in B20). This is the simplest measure of gravity wave dissipation that applies if the waves are fully damped -- regardless of the specific damping mechanism. Whether or not this is valid is an open question. This estimate is the simplest one to estimate the effects of gravity waves and was the one adopted in many prior works \citep[e.g.][]{Lubow1997,Ogilvie1} but future work should explore in detail the validity of this assumption in thin envelopes. We omit the influence of Coriolis forces here, partly for simplicity, and partly because the typical level of dissipation due to gravity (or gravito-inertial) waves is unlikely to differ substantially in this fully damped regime \citep[e.g.][though will likely differ to a greater extent for certain tidal frequencies involving resonances with inertial modes in neighbouring convective regions]{Ogilvie1,IvPapCh2013}.

When a planet possesses multiple radiative and convective envelopes, the total dissipation rates (not tidal quality factors) are derived by summing up the contribution from each layer where the dissipation takes place.

\subsection{Planetary model}
We compute planetary models using the {\sc MESA} code \citep[version r11701;][]{MESA1,MESA2,MESA3,MESA4,MESA5}. Most parameters in our \texttt{inlist} files are adopted from the \texttt{make\rule{0.15cm}{0.15mm}planets} test suit. We set \texttt{initial\rule{0.15cm}{0.15mm}Y} and \texttt{initial\rule{0.15cm}{0.15mm}Z} to 0.2804 and 0.02131, respectively, to reproduce the high average metallicity of hot Jupiter hosts (<[Fe/H]> = +0.19 dex, see \citealt{Petigura}). According to our exploration of parameter space, the tidal quality factors obtained with metallicities in the range between -0.5 and +0.5 dex are similar within an order of magnitude for any given age \citep[consistent with findings for the lowest mass stars considered in][]{Bolmont2}. Given that abundances do not seem to play a major role in any of our results, the effects of chemical composition will not be reported further in this paper. 
The planetary mass is varied between 0.1 and 10 $M_\mathrm{J}$, where $M_\mathrm{J}$ is the mass of Jupiter and we fix the initial radius to $2R_\mathrm{J}$ (increasing to $4R_\mathrm{J}$ for ``hot-start" models with higher initial entropy does produce substantial differences in $Q'$). The core mass is 10 $M_\mathrm{\oplus}$, where the subscript $\oplus$ refers to Earth units, and its density is 5 $\mathrm{g \; cm^{-3}}$, giving a fixed core radius of approximately $0.2R_\mathrm{J}$. Note that the core radius, when normalised by $R_\mathrm{pl}$, varies because the planetary radius (rather than the core size) evolves in time. The fiducial value of the incident flux is 1000 $F_\oplus$ and we use the canonical mixing length value $\alpha=2$, though we have explored smaller $\alpha$ values and found minimal differences in our results. 
The column depth for irradiation is fixed at 330 $ \rm g \; cm^{-2}$ to reproduce the mean opacity from \cite{Guillot}. The control \texttt{use\rule{0.15cm}{0.15mm}dedt\rule{0.15cm}{0.15mm}form\rule{0.15cm}{0.15mm}of\rule{0.15cm}{0.15mm}energy\rule{0.15cm}{0.15mm}eqn} is set to \texttt{.false.} to avoid convergence issues at late ages. We found that, for lower planetary masses, the default spatial resolution is too low to provide accurate solutions for the tidal response. Therefore, we choose a higher resolution by setting \texttt{max\rule{0.15cm}{0.15mm}dq = 1d-3} if there are no convergence issues at the beginning of the MESA run. Otherwise, the parameter \texttt{max\rule{0.15cm}{0.15mm}dq} is gradually increased until convergence issues are avoided.

Our planetary models generate a neutrally (adiabatically) stratified interior, as we might expect in convective regions, with only the surface layers being stably stratified (radiative). However, observational inferences from the Solar System's gas giants suggest this assumption may not be valid due to interior compositional gradients, and Jupiter or Saturn could possess extended dilute stably stratified fluid cores \citep[e.g.][]{Mankovich,Howard}. The consequences of interior stably stratified layers are outside the scope of the present paper, and are currently a major uncertainty in planetary models \citep[but see e.g.][]{Pontin2023,Lin2023,Dewberry2023,Dhouib2023,Pontin2023b}.

\section{Results}

\label{sec:results}
\subsection{Evolution of tidal quality factors} \label{evol}

\begin{figure*}
\begin{multicols}{2}
    \includegraphics[width=\linewidth]{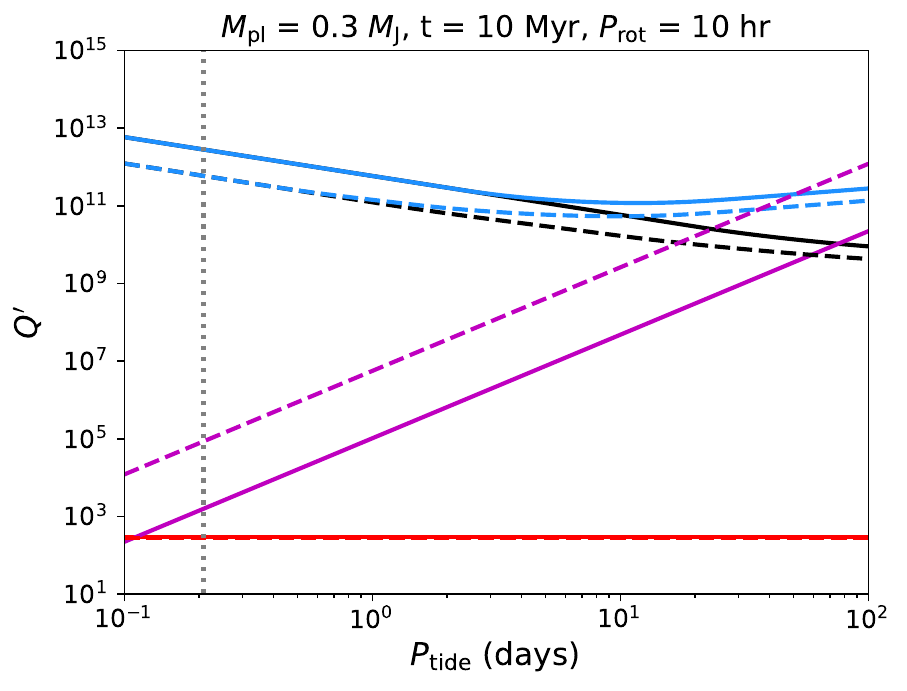}\par 
    \includegraphics[width=\linewidth]{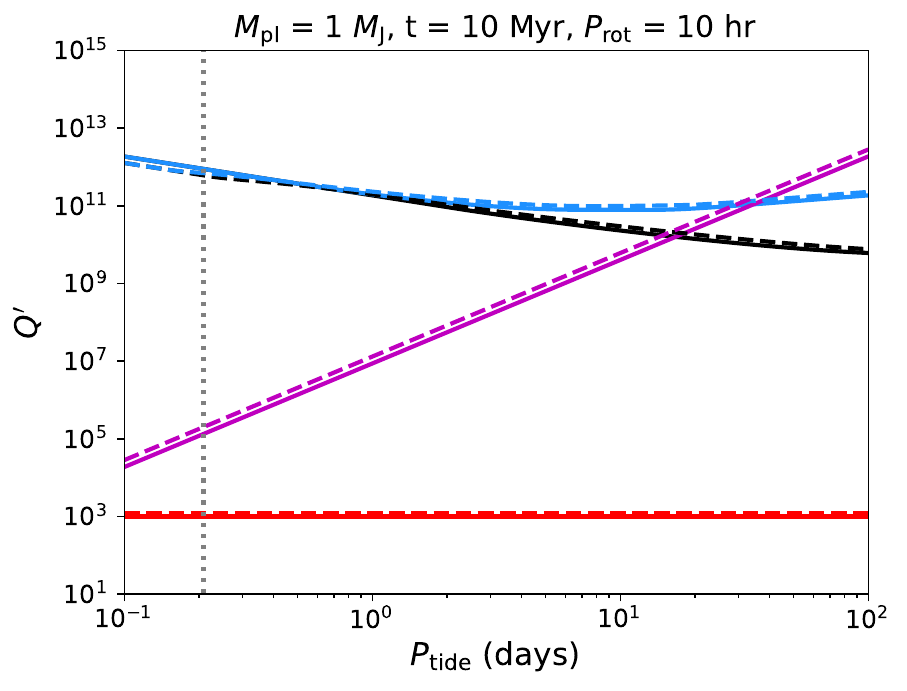}\par
    \includegraphics[width=\linewidth]{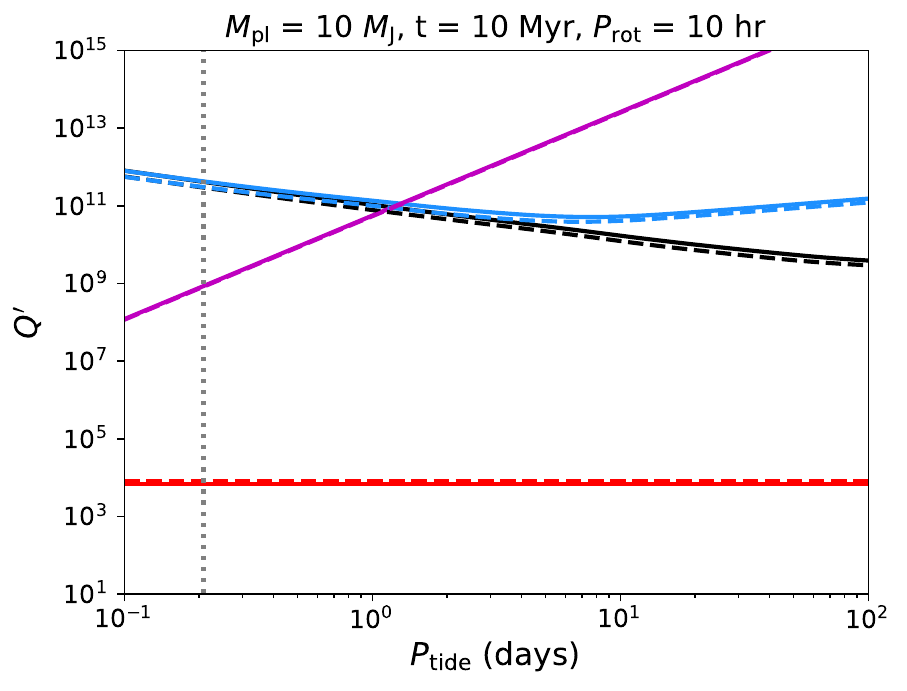}\par 
    \includegraphics[width=\linewidth]{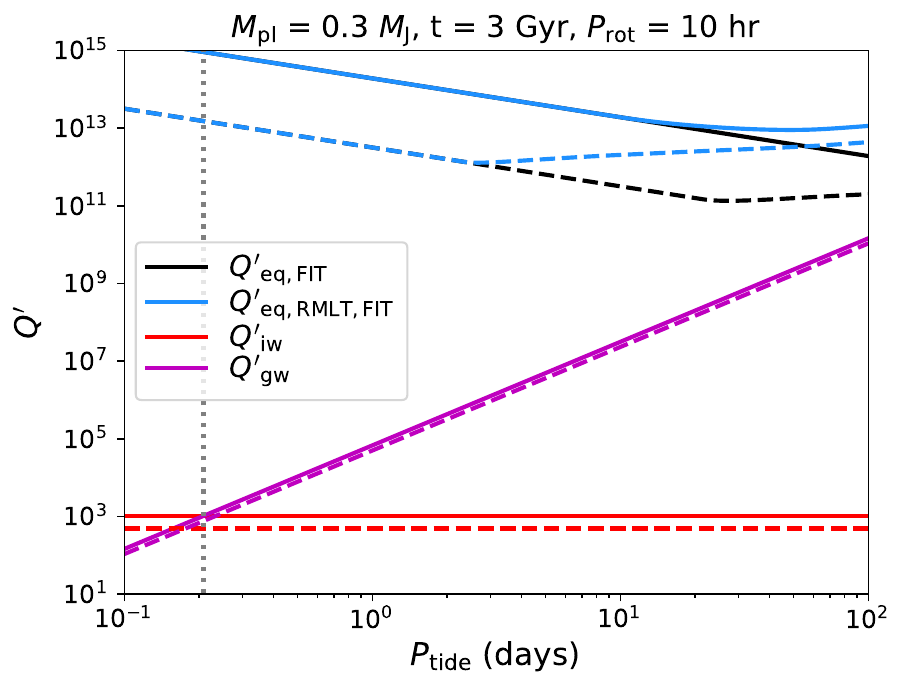}\par
   \includegraphics[width=\linewidth]{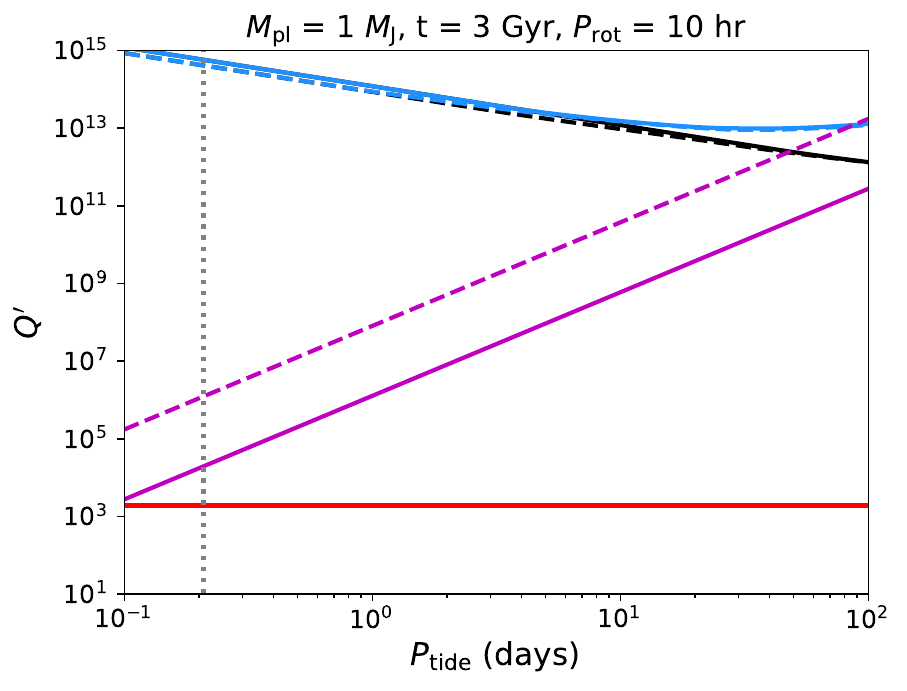}\par
    \includegraphics[width=\linewidth]{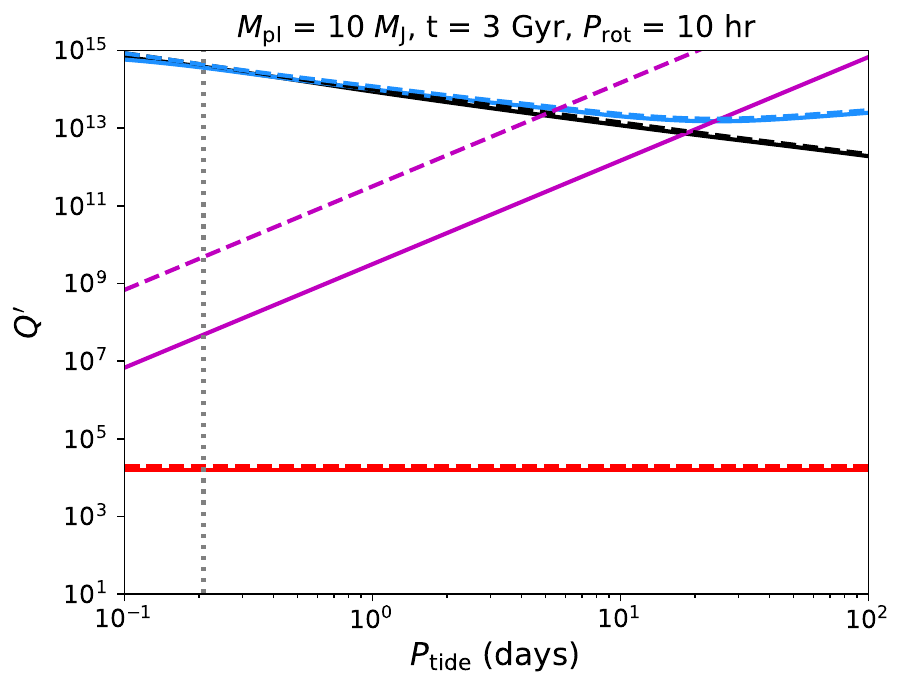}\par
\end{multicols}
    \vspace*{-6mm}
\caption{Tidal quality factors $Q'$ for each mechanism as a function of tidal period $P_\mathrm{tide}$. Solid and dashed lines correspond to the hot ($F$ = 1000 $F_{\oplus}$) and cold ($F$ = $F_{\oplus}$) models, respectively. The planetary mass, rotation period (fixed here to $P_\mathrm{rot}=10$ hr), and age are shown on the top of each plot. Black and blue lines represent tidal quality factors due to equilibrium tide damping without and with rotational modification, respectively. Red and magenta lines display tidal quality factors due to inertial and gravity waves, respectively. Grey dotted line illustrates the minimal tidal period required for inertial wave excitation ($P_\mathrm{tide} = P_\mathrm{rot}/2$).}
\label{fig2}
\end{figure*}

We now turn to present our results for $Q'$ computed in planetary models. We begin by showing the dependence of the tidal quality factors for each mechanism on the tidal forcing period $P_\mathrm{tide}$ in Fig.~\ref{fig2} for a range of planetary masses, ages, and instellations. Here, the top, middle, and bottom rows show planets with $M_\mathrm{pl}$ = 0.3, 1.0, and 10 $M_\mathrm{J}$, respectively. The left column corresponds to models of young Jupiters ($t$ = 10 Myr), and the right column represents models of old Jupiters ($t$ = 3 Gyr). The fiducial hot planets ($F$ = 1000 $F_{\oplus}$) are shown with solid lines and cold planets ($F$ = $F_{\oplus}$) are shown with dashed lines. These panels show a wide range of tidal frequencies, and hence our results can be applied to model spin-orbit synchronisation, the dominant component driving orbital circularisation, and aspects of tidal obliquity evolution.

As reported previously, equilibrium tide dissipation is very weak in all cases displayed according to our assumptions outlined in \S~\ref{sec:method}, with the minimum $Q'_\mathrm{eq}\sim 10^9$. At low tidal periods, the rotationally-modified tidal quality factor $Q'_\mathrm{eq, RMLT, FIT}$, shown in blue, is the same as the one obtained based on non-rotating convection $Q'_\mathrm{eq, FIT}$ (black, where subscript FIT refers to a fit from numerical simulations). This is because, in the high-frequency regime, $\nu_\mathrm{E} \propto \nu_\mathrm{MLT}  \left(\frac{\omega_\mathrm{c}}{\omega_\mathrm{t}} \right)^2 \propto \frac{u_\mathrm{c}^3}{l_\mathrm{c}}\frac{1}{\omega_\mathrm{t}^2}$, and the adopted rotationally-induced scalings for the convective velocity and length scale counteract each other \citep{Nils2023}. In the fast tides regime (with or without rotational inhibition of convection) -- which is typically the most relevant one in giant planets \citep[see e.g.][]{Nils2023} -- we thus have $Q'_\mathrm{eq}\propto P_\mathrm{tide}^{-1}$ because $|\omega_\mathrm{t}|/Q'_\mathrm{eq} \propto D_\mathrm{v}\propto \omega_\mathrm{t}^2\nu_E\propto \omega_t^0$. Thus, the slowness of convective flows relative to tides leads to substantial reductions in equilibrium tide damping.

With our chosen rotation period of 10 hr adopted for illustration (results for different $P_\mathrm{rot}$ can be obtained simply by re-scaling since $Q'_\mathrm{iw}\propto P_\mathrm{rot}^2$, so if $P_\mathrm{rot}=1$ day, $Q'_\mathrm{iw}$ should be a factor of 5.76 larger), inertial wave dissipation is the dominant tidal mechanism ($Q'_\mathrm{iw}$ is smallest) over the full range of tidal periods considered, except for the `old' model of 0.3 $M_\mathrm{J}$ planet, for which gravity waves begin to prevail at low $P_\mathrm{tide}$ (i.e., $Q'_\mathrm{gw}<Q'_\mathrm{iw}$). Note that $Q'_\mathrm{iw}$ only strictly operates if the tidal period $P_{\mathrm{tide}}>P_{\mathrm{rot}}/2$, otherwise inertial waves are not (linearly) excited and we should not employ $Q'_\mathrm{iw}$ to model tidal evolution. This is independent of frequency because we have adopted the frequency-averaged measure here -- in reality, inertial wave dissipation is expected to be strongly frequency-dependent, though this value is thought to be a representative one for tidal modelling of planetary populations as discussed in \S~\ref{sec:method}. In this short tidal period regime (left region compared to the dotted line), $Q'_\mathrm{gw}$ should be used instead according to Fig.~\ref{fig2}.

\begin{figure}
	\includegraphics[width=\columnwidth]{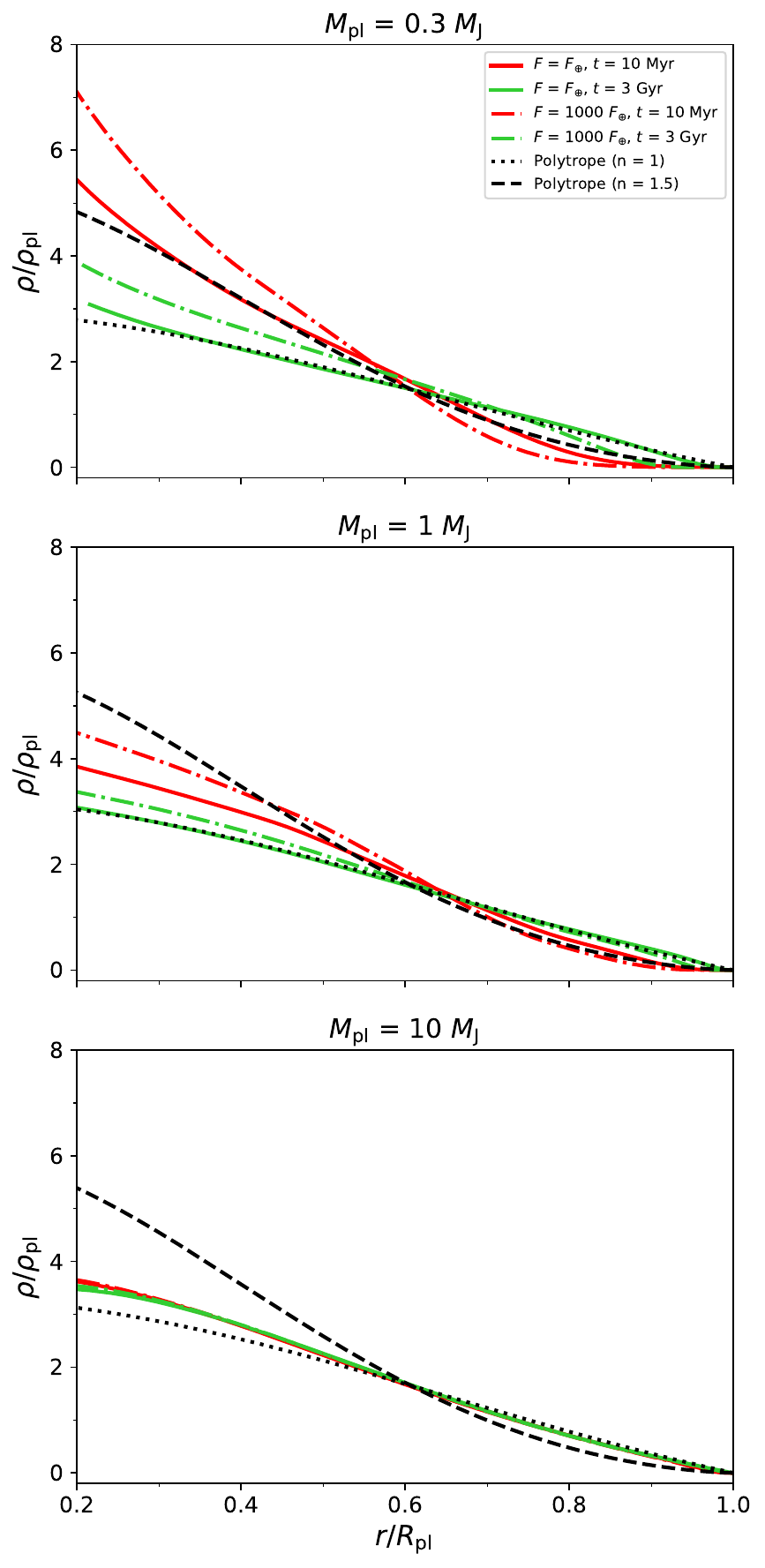}
    \vspace*{-6mm}
    \caption{Density profiles as a function of radius for the planetary gaseous envelopes of the models displayed in Fig.~\ref{fig2}. Density is normalised by the mean density, and radius is normalised by the planetary radius. Solid (dash-dotted) lines correspond to the `cold' (`hot') models; the red (green) color refers to the age of 10 Myr (3 Gyr). Black dotted and dashed lines correspond to the polytropic model with indices $n = 1$ and $n = 1.5$, respectively.}
    \label{fig5}
\end{figure}

The prediction for $Q'_\mathrm{iw}$ is similar in all models since they have the same rotation period and a similar internal structure. Indeed, each model has a structure for all ages that is very similar to a polytrope with a polytropic index ranging from $n=1$ (commonly thought to be appropriate for Jupiter) to $1.5$ (thought to apply to fully convective low-mass stars). As shown in Fig.~\ref{fig5}, where we plot density normalised by the mean density and radius normalised by the planetary radius, our models are well described by Lane-Emden polytropes with such a range of $n$, where $n=1$ and $n=1.5$ are represented by black dotted and dashed lines, respectively. The only exception is the models of young low-mass planets displayed in red in the top panel. Nevertheless, as these planets cool down with age, they approach the profile of the $n = 1$ polytrope. Accordingly, the models at 3 Gyr, depicted in green, yield flatter density profiles. In contrast to the `cold' models, shown with solid lines, highly-irradiated planets, represented by dash-dotted lines, are characterized by a steeper density gradient. At the same time, one can see that the internal structure of more massive planets, displayed in the bottom panel, is closer to the $n = 1$ polytrope and less sensitive to age and instellation. We, therefore, conclude that most of our models can be approximated by polytropic solutions with $n=1$ or $n=1.5$ with sufficient accuracy.

Adopting a polytropic model with index $n=1$ ($1.5$), we find $Q'_\mathrm{iw}=230.22 \,\omega_\mathrm{dyn}^2/\Omega_\mathrm{pl}^2$ (or $130.83\,\omega_\mathrm{dyn}^2/\Omega_\mathrm{pl}^2$), where $\omega_\mathrm{dyn}^2=G M_\mathrm{pl}/R_\mathrm{pl}^3$ is the squared dynamical frequency, implying a value approximately 2558 (1454) for a Jupiter-like model/rotation, similar to the values shown in Fig.~\ref{fig2}. Hence $Q'_\mathrm{iw}$ (for a fixed $P_\mathrm{rot}$) varies only modestly with planetary mass, age, and instellation within the ranges we consider, ultimately because planetary structures always remain very similar (and similar to polytropes) in our models.

Internal gravity waves become more dissipative (smaller $Q'_\mathrm{gw}$) in planets with thicker radiative envelopes, typically corresponding to higher stellar instellations (solid purple lines), lower planetary masses, and older ages (where planets have had time to develop thicker envelopes). In all cases $Q'_\mathrm{gw}\propto P_\mathrm{tide}^{8/3}$ under the assumptions of our model, and thus shorter tidal periods imply more efficient dissipation.

\begin{figure}
	\includegraphics[width=\columnwidth]{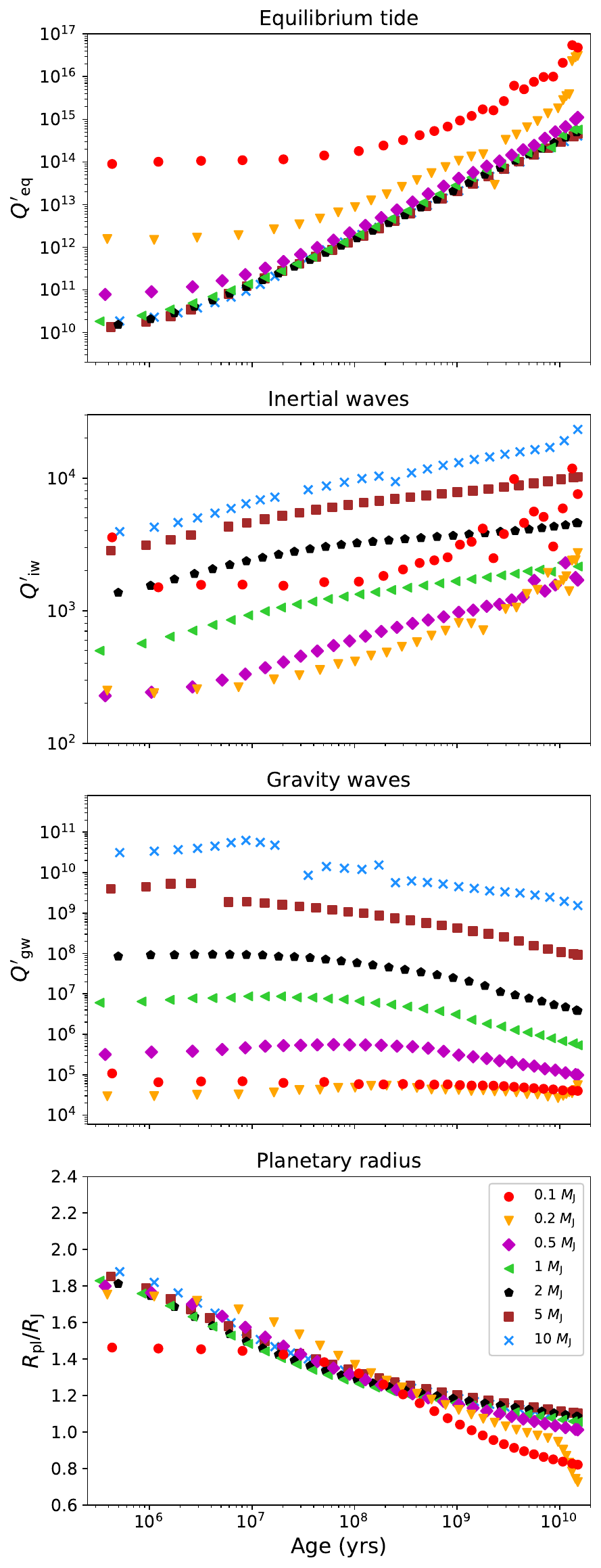}
    \vspace*{-6mm}
    \caption{Evolution of tidal quality factors with planetary age in a range of models due to equilibrium tides (first panel), inertial waves (second panel), and gravity waves (third panel). In the fourth panel, the evolution of the planetary radius is shown. Here, we fix tidal period $P_\mathrm{tide}$ = 1 day, rotation period $P_\mathrm{rot}$ = 10 hr, and incident flux $F$ = 1000 $F_{\oplus}$ (i.e.~these are `hot' planets).}
    \label{fig1}
\end{figure}

\begin{figure*}
	\includegraphics[width=\linewidth]{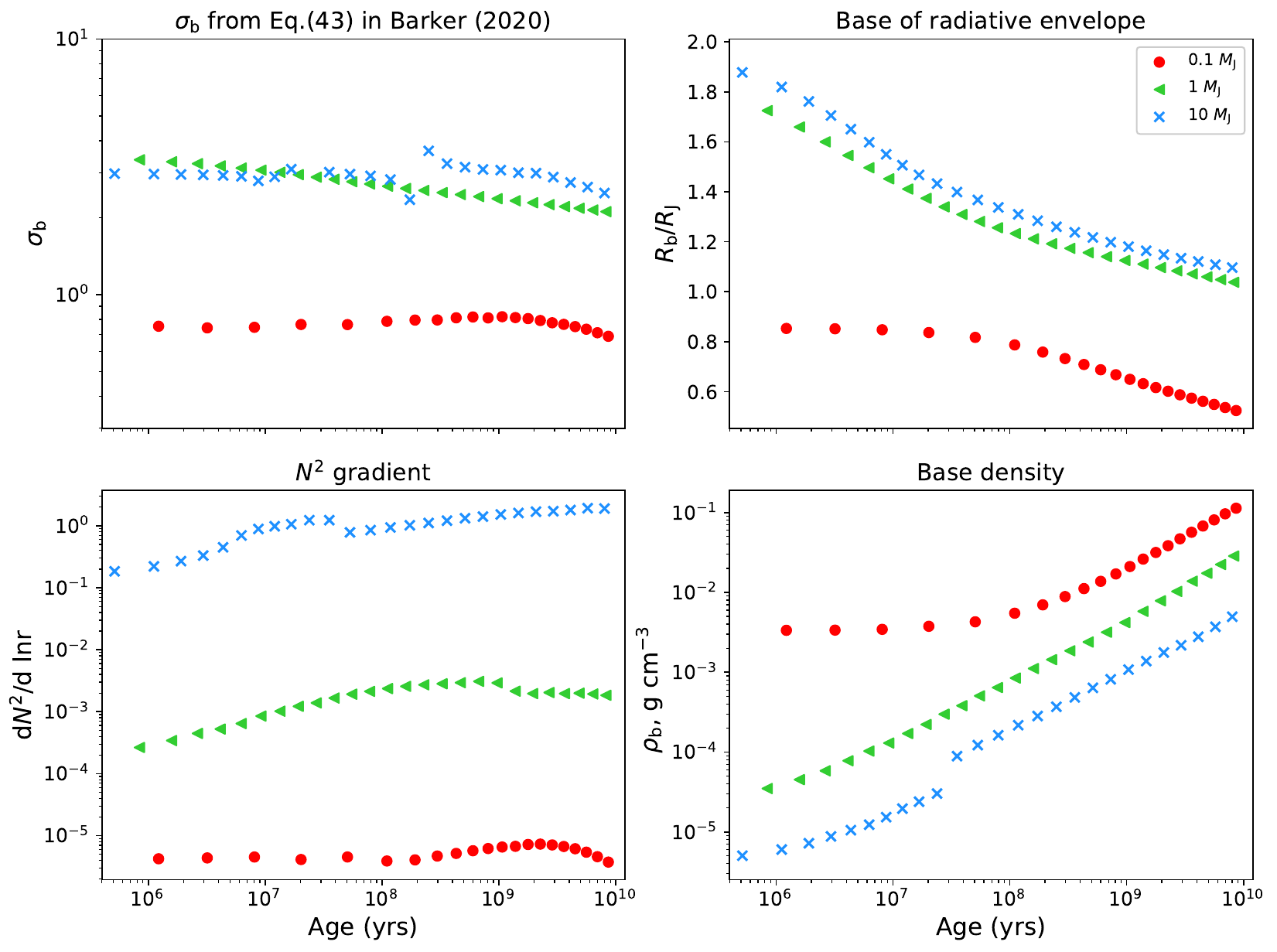}
    \vspace*{-6mm}
    \caption{Evolution of various quantities that are important in computing tidal quality factor due to gravity waves $Q'_\mathrm{gw}$ according to (\ref{eq:tide_gw1}) and (\ref{eq:tide_gw2}) in hot planetary models with $P_\mathrm{tide}$ = 1 day.}
    \label{fig6}
\end{figure*}
In Fig.~\ref{fig1}, we show the evolution with planetary age of the tidal quality factors corresponding to equilibrium tides (first panel), inertial waves (second panel), and gravity waves (third panel) for a set of `hot' planetary models with different masses. We now fix both the tidal and rotation periods at $P_\mathrm{tide}=1$ day and $P_\mathrm{rot}=10$ hr, respectively, to focus on the age and mass dependence here. The choice of $P_\mathrm{rot}=10$ hr is made for comparison with Jupiter, but we note that for inertial waves, we predict $Q'_\mathrm{iw}\propto P_\mathrm{rot}^2$ in general, so our results can be easily scaled for different rotation rates.

According to the prescriptions we have adopted, the equilibrium tide is characterized by negligible damping inside the convective envelope (using $Q'_\mathrm{eq, RMLT, FIT}$), insufficient to cause a significant change in orbital or spin parameters. The relevant tidal quality factor $Q'_\mathrm{eq}>10^{10}$ throughout the evolution of each of these planets. This is similar to the conclusions in B20 regarding the inefficiency of equilibrium tide dissipation in stellar interiors. Note that $Q'_\mathrm{eq}$ increases with age because the planet cools, thus convection slows down as the planet evolves, and because the planet shrinks.

In contrast, dissipation of inertial waves appears to be the most important mechanism in almost all cases when they are excited, with $Q'_\mathrm{iw}$ (for this $P_\mathrm{rot}$) ranging between $10^2$ and $2 \times 10^4$, with higher values corresponding to higher-mass objects. As shown in the second panel of Fig.~\ref{fig1}, planets gradually become less dissipative with age. We have explored the reason for this, and found that it is primarily not related to structural changes (consistent with Fig.~4 of B20 for low mass fully convective objects), but is instead explained by the shrinking radius as the planet cools for a fixed $P_\mathrm{rot}$ because $Q'_\mathrm{iw}\propto \omega_\mathrm{dyn}^2/\Omega_\mathrm{pl}^2\propto R_\mathrm{pl}^{-3}$. The evolution of planetary radius is displayed in the bottom panel. Furthermore, given that planetary spin-down is typically the natural outcome of long-term tidal and planetary evolution (unless, e.g., the planet is spiralling into its star while remaining tidally locked), we expect inertial wave damping to become less efficient at later epochs. 

On the other hand, the tidal quality factor due to gravity waves $Q'_\mathrm{gw}$ does not exhibit substantial evolution during the planetary lifetime, and its variation for each planetary model is within approximately an order of magnitude for a fixed $P_\mathrm{tide}$ and planetary mass. In addition to its strong dependence on $P_\mathrm{tide}$ ($Q'_\mathrm{gw}\propto P_\mathrm{tide}^{8/3}$), gravity wave damping also strongly depends on the planetary mass, spanning over six orders of magnitude for our mass range characteristic of gas giants. This is illustrated in the third panel of Fig.~\ref{fig1}. Similar to inertial waves, gravity waves dissipate more efficiently (smaller $Q'_\mathrm{gw}$) in lower-mass objects, with values as small as $Q'_\mathrm{gw}\approx 10^4$, whereas the most massive objects we consider are much less dissipative. These values are sensitive to the radius $r_\mathrm{b}$ and density $\rho_\mathrm{b}$ at the launching region, which is manifested through the factor $\mathcal{G}$ being proportional to $r_\mathrm{b}^5$ (Eq.~\ref{eq:tide_gw2}).

In Fig.~\ref{fig6} we explore in more detail the reasons for the substantial variation of $Q'_\mathrm{gw}$ with planetary mass in our models. Here, we illustrate the evolution of the quantities involved in the expression for $Q'_\mathrm{gw}$, given by Eqs. (\ref{eq:tide_gw1}) and (\ref{eq:tide_gw2}), for the models with $M_\mathrm{pl}$ = 0.1, 1, and 10 $M_\mathrm{J}$, depicted in red, green, and blue, respectively. The top left and bottom left panels display $\big|\d \,\mathcal{N}^2/\d \,\ln\,r\big|_{r=r_\mathrm{b}}$ and  $\sigma_\mathrm{b}$, while the top right and bottom right panels display $r_\mathrm{b}$ and $\rho_\mathrm{b}$ as a function of age, respectively. The value of $Q'_\mathrm{gw}$ obtained for planets with $M_\mathrm{pl} = 10 M_\mathrm{J}$ is, on average, 3.5 orders of a magnitude higher than for Jupiter-mass planets. These two planets are characterized by similar values of $\sigma_\mathrm{b}$ and $r_\mathrm{b}$, but there is an order of magnitude difference attributable to $\big|\d \,\mathcal{N}^2/\d \,\ln\,r\big|_{r=r_\mathrm{b}}$ (note that this quantity is raised to the power -1/3 in Eq.(\ref{eq:tide_gw1})), and half an order of magnitude from the variation in $\rho_\mathrm{b}$. The remaining factor of $10^2$ results from the presence of $M_\mathrm{pl}^2$ in Eq. (\ref{eq:tide_gw1}). At the same time, decreasing the planetary mass from 1 to 0.1 $M_\mathrm{J}$ reduces $Q'_\mathrm{gw}$ by a factor of $\sim 50-10^2$. A factor of $\sim 4$ arises due to differences in $\big|\d \,\mathcal{N}^2/\d \,\ln\,r\big|_{r=r_\mathrm{b}}$. One can see that $\sigma_\mathrm{b}$ is substantially smaller in the case of a lower-mass planet, leading to a factor of $\sim 3$ in $Q'_\mathrm{gw}$. An additional factor of $\sim 4$ (at t $\sim 1$ Gyr) comes from differences in $\rho_\mathrm{b}$, and one of $\sim 1.5$ arises from the combination $r_\mathrm{b}^5 R_\mathrm{pl}/M_\mathrm{pl}^2$. Combining these (crude) factors results in a total reduction of $\sim 50-10^2$ from 1 to 0.1 $M_\mathrm{J}$, in agreement with the overall differences outlined above. Therefore, we conclude that differences in several parameters come into play to produce the strong dependence on planetary mass exhibited by $Q'_\mathrm{gw}$ in our models, rather than any one of these parameters.

\subsection{Impact of the incident flux} \label{irrad}

\begin{figure}
	\includegraphics[width=\columnwidth]{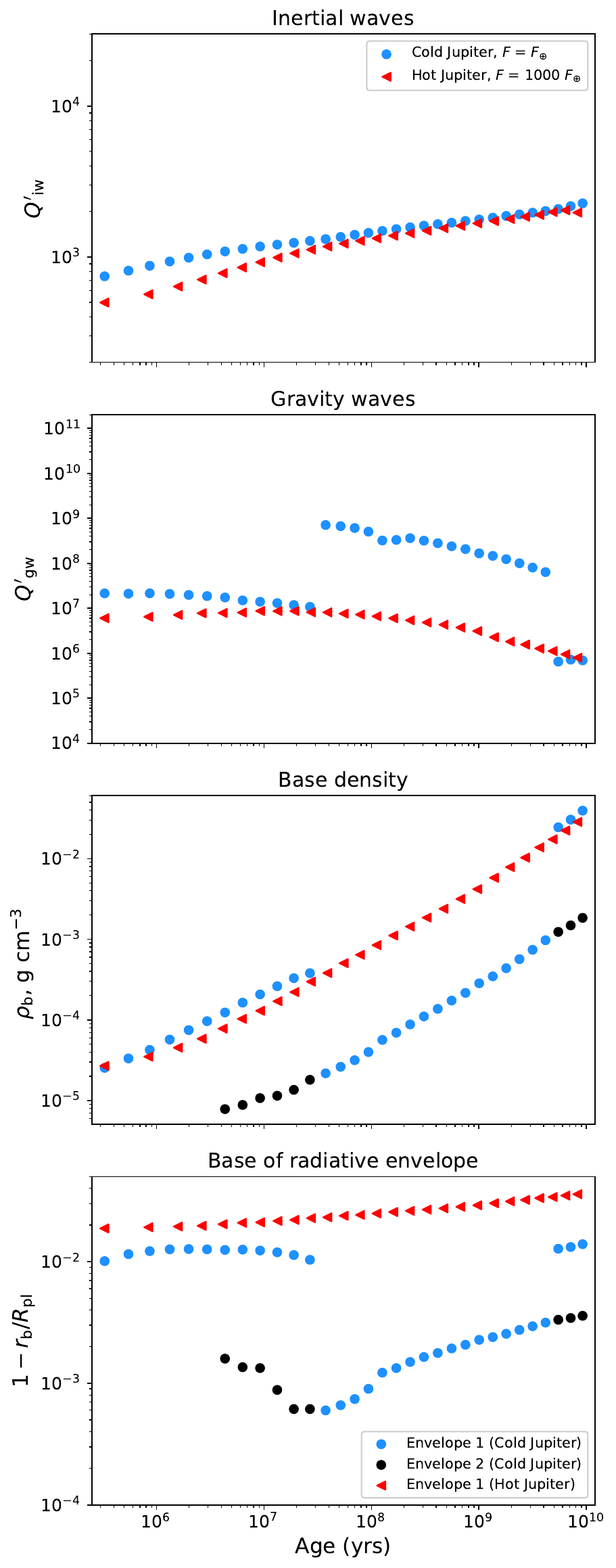}
    \caption{Evolution of $Q'$ due to inertial waves (top panel) and gravity waves (middle panel) for a Jupiter-mass planet irradiated by low ($F$ = $F_{\oplus}$) and high ($F$ = 1000 $F_{\oplus}$) incident fluxes (blue and red colors, respectively). Bottom panel: radius of the base of each radiative envelope with respect to the planetary surface. Blue and black circles represent the base of the bottom and top envelopes of a cold Jupiter, respectively. Red triangles represent the only radiative envelope of a hot Jupiter.}
    \label{fig3}
\end{figure}

The strength of external irradiation affects the locations of the interfaces between convective and radiative regions near the surface, thereby altering the conditions inside the layers where gravity waves are excited and propagate. This is demonstrated with the example of a Jupiter-mass planet in Fig.~\ref{fig3}. The top and middle panels display the evolution of the tidal quality factors of inertial and gravity waves, respectively. In the following plots, blue circles correspond to the low incident flux, characteristic of a `cold' Jupiter ($F$ = $F_{\oplus}$), and the red triangles represent a typical `hot' Jupiter irradiation ($F$ = 1000 $F_{\oplus}$). One can see that the enhancement of stellar flux received by a planet slightly increases the inertial wave dissipation rate, which is most noticeable at early ages, which is primarily because these planets have slightly inflated radii (see discussion above). Nonetheless, differences between the two models are always within an order of magnitude. On the contrary, the incident flux plays a crucial role in modifying gravity wave damping for most of the planetary lifetime. According to our `cold' model, $Q'_\mathrm{gw}$ rises by almost two orders of magnitude up to $10^9$ after 30 Myrs. Eventually, the dissipation rates are amplified to the values obtained for hot Jupiters after 6 Gyrs. 

As we show in the bottom panel of Fig.~\ref{fig3}, these dramatic changes in $Q'_\mathrm{gw}$ are directly linked to the location and number of radiative zones near the surface that arise in these models. Here, we display the separation of the base of each radiative zone from the planetary surface. Note that the lower the point is, the closer the corresponding interface is to the planetary surface. The prominent feature in the cold planetary model is the emergence of a second radiative region after 3 Myr. Thus, the bottom base is depicted by the blue circles, while the top base, when present, is illustrated by the black circles. As a result of the above, both envelopes could contribute to the dissipation of gravity waves. The overall tidal quality factor due to gravity waves is computed (crudely) by summing up the dissipation rates associated with each radiative region. As we mentioned previously, the efficiency of gravity wave damping is determined by the conditions at their launching sites, including the local density, characterized by a steep gradient in the vicinity of the planetary surface. Thereby, the occurrence of the outer radiative layer does not significantly alter the evolution of the tidal quality factor as long as the bottom layer exists. However, at the planetary age of 30 Myrs, two convective envelopes merge into one, leaving the top radiative region as the only contributor to gravity wave damping, which immediately manifests in a sharp increase in $Q'_\mathrm{gw}$, as shown in the middle panel. Finally, after 6 Gyrs, the radiative envelope becomes divided into two separate regions again, allowing the dissipation rates obtained for cold and hot Jupiters to converge. We have found that the appearance of multiple radiative layers is insensitive to \texttt{initial-Y} and \texttt{initial-Z} in the ranges $[0.25,0.28]$ and $[0.004,0.03]$, respectively, and the resulting values of $Q'_\mathrm{gw}$ do not differ substantially. It is possible that the emergence of these layers would differ with different equations of state to those used in our version of {\sc MESA}.

The comparison between models with different incident fluxes was also provided in Fig.~\ref{fig2}, where solid and dashed lines represent hot and cold Jupiters, respectively. Contrary to the earlier epoch, at late ages, planets with $M_\mathrm{pl}$ = 1 and 10 $M_\mathrm{J}$ reveal substantial variation in gravity wave dissipation rate with the amount of irradiation, with hot Jupiters being more dissipative. For 0.3 $M_\mathrm{J}$ planets, however, the changes in $Q'_\mathrm{gw}$ between cold and hot models are manifested at early times. In addition, for lower-mass planets, tidal quality factors due to equilibrium tides are also sensitive to the incident flux. In contrast to gravity waves, the dissipation of equilibrium tides is somewhat more effective in such low mass cold planets, which have smaller $Q'_\mathrm{eq}$.

We have assumed gravity waves are launched in each layer and are then fully damped before returning to their launching sites, even when there multiple radiative regions. It is uncertain whether this is justified, as is the emergence of a second radiative region, which may either be the natural outcome of planetary evolution for low incident fluxes, or it could be an artifact caused by uncertainties in the equation of state -- or various neglected physics -- in current versions of {\sc MESA}. We have discovered that this feature can be eliminated with the introduction of additional interior heating, which may arise due to tidal heating or Ohmic dissipation. The fully damped assumption can potentially be justified by linear radiative damping, though this might only be effective in the outer zone, but there are alternative possibilities (including differential rotation, non-linearity, and magnetic fields). These should be explored further in future work, but for now we caution that there are potentially large uncertainties in $Q'_\mathrm{gw}$, particularly in our cold models.

To summarise, we find gravity wave damping can be effective for highly-irradiated planets with extended stable layers near their surfaces that are deeper than one percent of their radii. Otherwise, inertial waves are predicted to be the most important tidal mechanism when they are excited (i.e.~for $P_\mathrm{tide}>P_\mathrm{rot}/2$) in almost all models, except perhaps for the latest ages for low masses when $Q'_\mathrm{iw}>Q'_\mathrm{gw}$.

\section{Application to star-planet and planet-moon systems}
\label{sec:discussion}

\begin{figure}
	\includegraphics[width=\columnwidth]{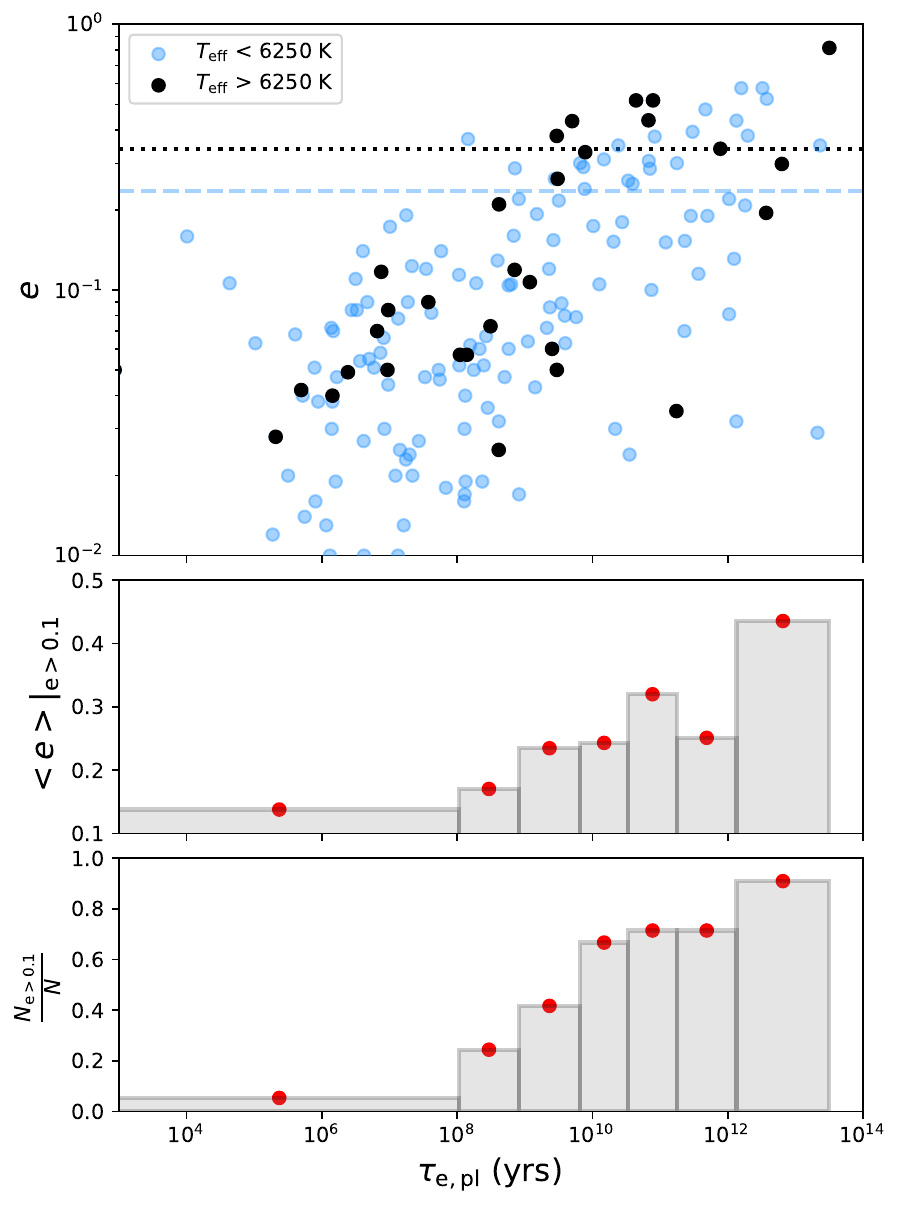}
    \caption{Top panel: eccentricity distribution for observed planets as a function of predicted tidal circularization timescale from planetary tides due to dissipation of inertial waves. Systems with the star above (below) the Kraft break are displayed in black (light blue). Blue dashed and black dotted lines illustrate the mean eccentricity among the planets with $e > 0.1$ (i.e., eccentric planets) orbiting stars below and above the Kraft break, respectively. Histogram in the middle panel shows the average eccentricity of the eccentric planets. Histogram in the bottom panel shows the ratio of the number $N_\mathrm{e>0.1}$ of eccentric planets to the number $N$ of planets within each bin.}
    \label{fig4}
\end{figure}
The dissipation of planetary tides can potentially explain an important aspect of the eccentricity distribution in star-planet systems, which is that hot Jupiters tend to have smaller eccentricities (and a stronger preference for circularity) than warm and cold Jupiters. To explore this scenario, we collect data from the NASA Exoplanet Archive (\url{https://exoplanetarchive.ipac.caltech.edu/}) representing massive close-in planets ($0.1\;M_\mathrm{J} < M_\mathrm{pl} < 10 \;M_\mathrm{J}$, $P_\mathrm{orb} < 20\;{\rm days}$; $P_\mathrm{orb}$ is the orbital period). In addition, we filter out systems containing another planet with $P_\mathrm{orb} < 100\;{\rm days}$ to avoid scenarios where eccentricity excitation due to planet-planet interactions may be competing with tidal eccentricity damping. Our main sample consists of 162 systems with a known eccentricity, stellar effective temperature, stellar and planetary mass, and planetary radius. This sample has been further extended by eight systems, namely HAT-P-2, HAT-P-4, HD 118203, HD 149026, HD 189733, HD 209458, Kepler-91, and WASP-8, with no accessible data on the planetary radii. The radii of the corresponding planets have been derived using the mass-radius-flux parametrization from \cite{Lazovik1} given by his Eqs. (16)—(18), which encompasses the relations from \cite{Valsecchi} and \cite{Thorngren}. We also consider 136 additional planets with an upper bound on the eccentricity below 0.1. We assume that these planets are on circular orbits ($e=0$) when calculating the relative number of eccentric planets (i.e., the planets with $e>0.1$).

For every system in our sample, we calculate a corresponding circularization timescale due to planetary tides, $\tau_\mathrm{e,pl}$, following the equation \citep[e.g.][]{GS1966}:
\begin{equation}
\tau_\mathrm{e,pl} = \frac{4}{63}\frac{Q'_\mathrm{pl}}{n}\frac{M_\mathrm{pl}}{M_\mathrm{*}}\left(\frac{a}{R_\mathrm{pl}} \right)^5.
    \label{eq:tau}
\end{equation}
Here, the tidal quality factor $Q'_\mathrm{pl}$  is set equal to $Q'_\mathrm{iw}$ since, as shown in Sec.~\ref{sec:results}, inertial waves provide the main contribution to the overall tidal dissipation inside the planets in almost all cases in our models. $Q'_\mathrm{iw}$ may be represented as the product of two components, namely the structural tidal quality factor $Q'_\mathrm{iw,s}$ and the parameter $\epsilon_\mathrm{\Omega}^{-2} \equiv \left(\Omega_\mathrm{pl} / \sqrt{GM_\mathrm{pl} / R_\mathrm{pl}^3} \right)^{-2}$. The first component, $Q'_\mathrm{iw,s}$, is computed via linear interpolation between our models of hot planets with the adjacent masses and ages plotted in Fig.~\ref{fig1} (note the large observational uncertainties in ages does not lead to large differences according to this figure), while $\epsilon_\mathrm{\Omega}$ is inferred using observational data, assuming spin-orbit synchronization ($\Omega_\mathrm{pl} = n$, since this is expected to occur more rapidly than circularization). 

Our sample is illustrated in the top panel of Fig.~\ref{fig4}. One can see that eccentricities tend to increase with the predicted circularization timescale. This is especially true for planets orbiting stars above the Kraft break ($T_\mathrm{eff}$ > 6250 K, \citealt{Kraft}), as shown in black. Hot stars have thin convective envelopes, leading to weaker tidal dissipation in stellar interiors (see B20), suggesting planetary tides may be even more important for eccentricity evolution in these systems. This weaker stellar tidal dissipation may have several observational manifestations. In particular, star-planet systems with a hot star typically sustain higher obliquities, as shown in \cite{Spalding, Attia}. Here, we draw similar conclusions concerning the eccentricity distribution, which reveals the same trend, albeit with caution due to the low numbers involved. Indeed, among the systems with $e> 0.1$, planets orbiting stars above (below) the Kraft break have an average eccentricity of 0.33 (0.24). Apart from this trend, a correlation between eccentricity and eccentricity damping timescale appears to be more pronounced in hot stars, which may imply that stellar tides also contribute to the orbital circularization of hot Jupiters, or it could be related to the shorter main-sequence ages of hotter stars.

As reviewed in \cite{DawsonJohnson}, hot Jupiters might have formed via two channels, namely disc migration and high-eccentricity migration (e.g.~triggered by planet-planet scattering or secular/Kozai migration). In contrast to the former channel, the latter allows the formation of highly eccentric hot planets. It is still unknown which channel (if any) dominates within the overall hot Jupiter population. The presence of almost circular systems with $\tau_\mathrm{e,pl}$ a few orders of magnitude higher than the age of the universe, seen in Fig.~\ref{fig4}, suggests that disc migration/low eccentricity formation is likely to be a favorable scenario for some fraction of our sample. To avoid the planets which might have formed with low initial eccentricities, we separate the planets with $e > 0.1$ (``eccentric planets") and $e < 0.1$ (``non-eccentric planets"). We select bin sizes along the $x$-axis to have roughly equal numbers of eccentric planets in each one. For every bin, we calculate the average eccentricity of the eccentric planets and plot it in the middle panel of Fig.~\ref{fig4}. In addition, we derive the fraction of planets per bin with $e > 0.1$, and we display this in the bottom panel. Both quantities are found to increase with our predicted tidal circularization timescale. There is only a handful of eccentric planets with $\tau_\mathrm{e,pl} < 10^8$ yrs. This might be expected if planetary tides have acted here given that the average age of the observed systems is on the order of a few Gyr. Another prominent detail is that the mean eccentricity of the eccentric sub-sample increases when $\tau_\mathrm{e,pl} \sim 1$ Gyr, i.e., when $\tau_\mathrm{e,pl}$ becomes comparable with the systems' mean age. The above features strongly suggest that tidal dissipation due to inertial waves can play an important role in shaping the orbital architectures of star-planet systems containing giant planets. On the other hand, under the assumptions of our models, neither equilibrium tides nor gravity waves can explain tidal circularisation timescales consistent with observations (that are shorter than or comparable to the ages of the systems). Gravity waves could circularise only the very closest planets under the assumptions we have made to model them \citep[neglecting the possibility of resonance locking, e.g.][]{Fuller2016}.

From the detailed analysis of astrometric observations of Jupiter's and Saturn's satellites, \cite{Lainey2009,Lainey2017} constrained tidal dissipation rates in our Solar System's giants. They inferred $k_2/Q = (1.102 \pm 0.203) \times 10^{-5}$ for Jupiter and $k_2/Q = (1.59 \pm 0.54) \times 10^{-4}$ for Saturn, giving approximately $Q' = (1.59 \pm 0.25) \times 10^5$ and $Q' = (9.43 \pm 4.39) \times 10^3$ for these planets, respectively. According to our models for evolved planets at the Solar System's age (Figs.~\ref{fig2} and ~\ref{fig1}), we find $10^3 \lesssim Q'_\mathrm{iw}\lesssim 10^4$ for $P_\mathrm{rot}\approx 10$ hr. Hence, inertial waves are sufficiently dissipative -- according to the frequency-averaged measure we have computed -- to explain observations. Given that the actual dissipation rate, and hence $Q'$ value, due to inertial waves at a given tidal frequency can vary by orders of magnitude from this ``typical value" represented by the frequency-average \citep[e.g.][in both linear and nonlinear calculations]{O2013,AB2022}, this suggests that the orbital evolution of Jupiter's and Saturn's moons can be explained by inertial waves \citep[see also][]{Pontin2022,Lin2023,Dewberry2023,Nils2023,Pontin2023b,Dhouib2023}. However further work is required to explore this scenario in more detail, and to determine the validity of the frequency-averaged formalism in modelling tidal evolution.

According to Fig.~\ref{fig1}, the above constraints on Jupiter and Saturn can also be obtained via gravity wave damping in the envelope. For the rotation period of Saturn ($P_\mathrm{rot} = 0.44$ days) and the orbital period of Enceladus ($P_\mathrm{orb} = 1.37$ days), our Saturn model predicts $Q'_\mathrm{gw} \sim 3 \times 10^3$ (not strongly depending on instellation for the relevant mass and age). In turn, the present-epoch gravity wave dissipation rate calculated for the Jupiter model strongly depends on the incident flux. Adopting $P_\mathrm{rot} = 0.41$ and $P_\mathrm{orb} = 1.77$ days (Jupiter's rotation period and Io's orbital period, respectively) yields $Q'_\mathrm{gw} \sim 4 \times 10^4$ for a hot Jupiter model and $Q'_\mathrm{gw} \sim 2 \times 10^6$ for a cold model. Thus, our crude gravity wave dissipation estimates are also (surprisingly) in reasonable agreement with the observations. It would be interesting to explore the role of interior stably-stratified layers in future work, which we have neglected in our models due to the large uncertainties involved \citep[i.e., to build upon][]{Fuller2016,Andre2017,Andre2019,Pontin2022,Pontin2023,Pontin2023b,Lin2023,Dewberry2023,Dhouib2023}.

\section{Conclusions}
\label{Conclusions}

We have studied theoretically the evolution of tidal dissipation rates and modified quality factors $Q'$ in rotating giant planets following their evolution using {\sc MESA} interior models with masses ranging from $0.1$ to $10M_\mathrm{J}$, for various incident stellar fluxes. We compute the dissipation of equilibrium tides by rotating turbulent convection (assuming an effective viscosity consistent with hydrodynamical simulations), dissipation of gravity waves in the thin radiative envelope, and inertial waves in the convective interior. 

Our models indicate that inertial waves are almost always likely to be the dominant mechanism of tidal dissipation in giant planets whenever they are excited\footnote{These waves can also be excited ``nonlinearly" by the elliptical instability for the hottest (very shortest period) planets, which we do not study here \citep[e.g.][]{Nils2023}.} -- i.e., when the tidal period $P_\mathrm{tide}>P_\mathrm{rot}/2$ -- and are capable of providing $Q'_\mathrm{iw}\sim 10^3 (P_\mathrm{rot}/10 \mathrm{hr})^2$. This implies $Q'_\mathrm{iw}\sim 10^5-10^6$ for orbital periods of order $10$ days (assuming spin-orbit synchronism). Note that the frequency-averaged measure we have adopted can differ by orders of magnitude from the predictions at a specific tidal frequency according to linear and nonlinear calculations \citep[e.g.][and can depend on magnetic fields and differential rotation]{O2013,LO2018,AB2022,AB2023}, but is likely to represent a rather robust ``typical value" of dissipation due to these waves.

In hot low-mass planets (approx $0.1M_\mathrm{J}$), our models also predict efficient dissipation of gravity waves in the radiative envelope with $Q'_\mathrm{gw}\sim 10^4 (P_\mathrm{tide}/1 \,\mathrm{d})^{8/3}$ \citep[see also][]{Lubow1997,Ogilvie1}. This indicates efficient dissipation via this mechanism is also possible, though $Q'_\mathrm{gw}$ values ranging up to six orders of magnitude larger are found in the more massive planets we modelled.

We have shown that our predicted circularization timescales from the dissipation of inertial waves correlate well with observed planetary eccentricities. This provides evidence that inertial wave dissipation may have played an important role in planetary tidal evolution.

The values of $Q'$ we have obtained can be compared with the latest statistical inferences from modelling exoplanetary eccentricity damping in \cite{Mahmud2023}. They found $Q'=10^{5\pm0.5}$ for hot Jupiters with $P_\mathrm{tide}\in[0.8,7]$ days, with no strong evidence of any tidal period dependence. For eccentricity tides, assuming spin-orbit synchronism (hence $P_\mathrm{tide}=P_\mathrm{orb}=P_\mathrm{rot}$), we predict $Q'_\mathrm{iw}\approx 10^{3.5}$ for $P_\mathrm{tide}=0.8$ days, and $ Q'_\mathrm{iw}\approx 10^{5.5}$ for $P_\mathrm{tide}=7$ days, with a value of $Q'_\mathrm{iw}\approx 10^{4.5}$ for $P_\mathrm{tide}=2.4$ days. Our results are therefore consistent with the range they obtained for tidal periods longer than about 2.4 days, and thus we argue that inertial waves are likely to be able to explain their results in these cases. For the shortest tidal periods, we find more effective dissipation than they do, though this may be mitigated when considering the frequency-dependent tidal dissipation. It is also unclear whether their assumption of a constant planetary radius affects their results.

Convective damping of equilibrium tides is estimated to be negligible in giant planets compared with wavelike tides because of the strong frequency-reduction of the effective turbulent viscosity due to the slow convection (relative to the tide) in these bodies \citep{GN1977,OL2012,DBJ2020b,Vidal}, -- though see \cite{T2021} for an alternative viewpoint. Rapid rotation minimally affects the resulting convective turbulent viscosities in the fast tides regime \citep[e.g.][]{Nils2023}, though it reduces the effective viscosity even further for slow tides. Further work should explore the interaction between tidal flows and convection in more realistic numerical models.

It is essential in future work to study whether the frequency-averaged formalism for inertial wave dissipation is appropriate to model tidal interactions when global inertial modes are excited in realistic density-stratified models of planets, and whether it faithfully reproduces overall trends resulting from the dynamical evolution in a population of planetary systems. The role of interior stably-stratified layers such as inferred for the dilute cores of Jupiter \& Saturn should also be explored further, as should the effects of differential rotation and magnetic fields.

\section*{Acknowledgements}
We would like to thank the initial referee, Caroline Terquem, for reading two versions of our manuscript and for providing critical comments each time, even if we disagree with many of them, and our second referee, Jim Fuller, for a constructive report that helped us to improve the paper. AJB was supported by STFC grants ST/S000275/1 and ST/W000873/1. NBV was supported by EPSRC studentship 2528559. AA was supported by a Leverhulme Trust Early Career Fellowship (ECF-2022-362).

\section*{Data Availability}
The data underlying this article will be shared on reasonable request to the corresponding author.



\bibliographystyle{mnras}
\bibliography{example} 




\bsp	
\label{lastpage}
\end{document}